# Use of Resonance Raman Spectroscopy to study the phase diagram of $PbZr_{0.52}Ti_{0.48}O_3$

J. Rouquette[1,2], J. Haines[1], V. Bornand[1], M. Pintard[1], Ph. Papet[1], J.L Sauvajol[3]

[1]Laboratoire de Physico-Chimie de la Matière Condensée, UMR CNRS 5617, Université Montpellier II, Place Eugène Bataillon, cc003, 34095 Montpellier cedex 5, France.

[2]Bayerisches Geoinstitut, Universität Bayreuth, D-95440 Bayreuth, Germany

[3]Laboratoire des Colloïdes, des Verres et des Nanomatériaux, UMR CNRS 5587, Université Montpellier II, cc026, Place E. Bataillon, F-34095 Montpellier, Cedex 5, France.

Abstract

Evidence is presented for the first time that the sharp and continuous spectral changes observed in $PbZr_{0.52}Ti_{0.48}O_3$ (PZT) between 350 and 10 K with the 647.1 nm wavelength are due to a resonance Raman effect. Such a phenomenon can be explained by means of a self trapped exciton emission oxygen deficient complex ($Ti_{Ti}'$-$V_O^{\cdot\cdot}$) of PZT powder whose energy is close to the radiation line of the laser. This kind of approach should also be very useful to distinguish the phase transition sequence for other related ferro/piezoelectric systems.

77.84.Dy, 77.80.Bh, 71.35.Aa.

# I. INTRODUCTION

Perovskite PbZr$_{1-x}$Ti$_x$O$_3$ ceramics are the most widely used ferroelectric materials because of their excellent properties. The highest piezoelectric and electromechanical coupling coefficients have been obtained for compositions near the morphotropic phase boundary (MPB)[1] between rhombohedral (*R3m*) and tetragonal (*P4mm*) phases (x ≈ 0.48). The MPB has generated much interest, especially since the discovery of a monoclinic phase[2] (*Cm*) in the MPB region for temperatures below 300 K. In this monoclinic form, the polarization can be modulated for example as a function of an external applied electric field between the rhombohedral and tetragonal directions based on a polarization rotation mechanism thus providing a possible explanation for this high piezoelectric response[3]. Additionally, a second monoclinic phase[4] with a doubled unit cell (*Cc*) has been observed at low temperature (T ≤ 210 K) for PbZr$_{0.52}$Ti$_{0.48}$O$_3$. This complex phase diagram around the MPB does not take into account the static disorder of the zirconium and titanium ion displacements[5,6] which introduces the well-known ferroelectric domain structure. Furthermore, recent experiments based on electron diffraction[5] and neutron diffraction[7] studies, Raman scattering[8] results, as well as theoretical calculations[9] have indicated the existence of an intrinsic short range dynamic disorder over nearly the entire PZT solid solution predominantly due to off-center lead displacements[7,10]. Therefore, the boundaries between the different phases, in particularly close to MPB compositions are still subject to debate. Based on Raman measurements[8,11-14] for compositions near the MPB, only minor spectral changes occurring in the temperature-composition range near the MPB were observed. This was associated with a two-phase coexistence[8] as has been the case for more than 30 years.

In this contribution, we investigate the different phase transitions of PbZr$_{0.52}$Ti$_{0.48}$O$_3$ (PZT) between 10 and 350 K by resonance Raman spectroscopy. Based on the temperature dependence in the frequencies of the vibrational modes, we clearly identified the transformations previously characterized by dielectric measurements[15-16], X-ray[17] and neutron diffraction[4]. The resonance phenomenon can be explained by the existence of a self-trapped exciton emission of an oxygen deficient complex of PZT powder (Ti$_{Ti}$'-V$_O^{..}$). Thus, resonance Raman scattering spectroscopy is one of the most useful methods to identify the phonon modes which couple to an electronic system[18-19]. As many perovskite oxides display very similar visible photoluminescence between 2 and 3 eV[20-21], Raman Resonance Spectroscopy of self-trapped exciton could be used in a similar way to clarify complex phase diagrams of MPB-ferroelectric or -relaxor systems.

## II. EXPERIMENTAL

PbZr$_{0.52}$Ti$_{0.48}$O$_3$ solid solutions was prepared by the conventional solid state reaction from high-purity (>99.9%) oxides via a two-stage calcination process[22-28]. A first set of ambient-temperature experiments was performed using a Jobin Yvon Labram spectrometer with a 632.8 nm He-Ne excitation line and laser output powers of 8 mW. The laser beam was focused with a 50× lens leading to a spot of approximately 2-3 μm of diameter. Additional variable-temperature Raman experiments (10-350K) were performed using an Oxford Instrument Microstat closed-cycle helium cryostat. The Raman spectra were obtained with an Argon-Krypton laser (488, 514.5 and 647.1 nm) and a Jobin-Yvon

T64000 triple monochromator equipped with an Olympus microscope and a CCD cooled to 140 K. A 50× objective was used to investigate sample regions with a diameter of about 3 µm. The laser output powers were 8 mW. The temperature was measured near the sample using a silicon diode.

### III. RESULTS AND DISCUSSION

In a previous study[28], Raman Spectroscopy with the 647.1 nm excitation line was used as a function of temperature to identify the boundaries between tetragonal and monoclinic (with simple and double-unit cell) ferroelectric phases previously observed in $PbZr_{0.52}Ti_{0.48}O_3$ (PZT) by dielectric measurements, X-ray and neutron diffraction. The obtained data were clearly different from the 514.5 nm Raman measurements generally reported in the literature[8,11-14], in which the temperature dependence of the broad vibrational modes was minor. The 647.1 nm $Kr^+$ line was used in the previous study[28] as PZT samples are known to absorb rather strongly in the high-energy visible range[30]. A carefully analysis of room temperature Raman spectra of $PbZr_{0.52}Ti_{0.48}O_3$ using several sets of excitation lines (488, 514.5, 632.8 and 647.1 nm), Fig. 1, reveals a more interesting phenomenon. On one hand, between 488 nm and 514.5 nm, Raman spectra are almost the same and the broad vibrational modes are at the same position. With respect to these previous results, significant changes are observed for the 632.8 nm and 647.1 nm spectra. For these latter spectra, the main features are an increase in intensity at low-frequency (150 - 300 $cm^{-1}$), and a change in the profile of the band located around 200 $cm^{-1}$. However, in the range above 170 $cm^{-1}$, the spectra obtained at 647.1 nm (1.92 eV) and 632.8 nm (1.96

5eV) are also different.The main differences occur in the region of the broad features located between 170-350 cm$^{-1}$ and 400-650 cm$^{-1}$. Especially, in the spectrum measured at 647.1 nm, one can note the appearance of a new and intense component at 510 cm$^{-1}$, of a weak line around 413 cm$^{-1}$ and the shifts of the two main components located in the 170-350 cm$^{-1}$ region. In order to explain this difference between the behavior using the high- (488 and 514.5 nm) and low- (632.8 nm and 647.1 nm) energy excitation lines on one hand, and between the 632.8 nm and 647.1 nm spectra on the other hand, one can link the 647.1 nm and 632.8 nm energies (1.92 and 1.96 eV) with the 2.0 eV luminescence observed by Liu et al. at ambient temperature for PbZr$_{0.52}$Ti$_{0.48}$O$_3$ films and ceramics[29]. Thus, the wavelength of the probe light may be more or less tuned to a resonance excitation of PZT sample. The maximum phonon-intensity in the resonance phenomenon is well-known to occur for energies slightly different from that of the photoluminescence line, which may explain the difference in behaviour for the 632.8 nm and 647.1 nm spectra (j'ai enlever smooth?). The photoluminescence at 2.0 eV was previously assigned by Trepakov et al. to a self trapped exciton emission of an oxygen deficient complex (Ti$_{Ti}$'-V$_O^{\cdot\cdot}$) for PLZT ceramics[30] and these authors have also proposed that this emission could be observed in other self-activated ATiO$_3$ perovskite-structure materials. Note that in our sample, oxygen vacancies can easily be explained at high temperature by the loss of PbO weight during sintering[22-28], therefore we have:5 eV) are also different. The main differences occur in the region of the broad features located between 170-350 cm$^{-1}$ and 400-650 cm$^{-1}$. Especially, in the spectrum measured at 647.1 nm, one can note the appearance of a new and intense component at 510 cm$^{-1}$, of a weak line around 413 cm$^{-1}$ and the shifts of the two main components located in the 170-350 cm$^{-1}$ region. In order to explain this difference between the behavior using the high- (488 and 514.5 nm) and low- (632.8 nm and 647.1 nm) energy excitation lines on one hand, and between the 632.8 nm and 647.1 nm spectra on the other hand, one can link the 647.1 nm and 632.8 nm energies (1.92 and 1.96 eV) with the 2.0 eV luminescence observed by Liu et al. at ambient temperature for PbZr$_{0.52}$Ti$_{0.48}$O$_3$ films and ceramics[29]. Thus, the wavelength of the probe light may be more or less tuned to a resonance excitation of PZT sample. The maximum phonon-intensity in the resonance phenomenon is well-known to occur for energies slightly different from that of the photoluminescence line, which may explain the difference in behaviour for the 632.8 nm and 647.1 nm spectra (j'ai enlever smooth?). The photoluminescence at 2.0 eV was previously assigned by Trepakov et al. to a self trapped exciton emission of an oxygen deficient complex (Ti$_{Ti}$'-V$_O^{\cdot\cdot}$) for PLZT ceramics[30] and these authors have also proposed that this emission could be observed in other self-activated ATiO$_3$ perovskite-structure materials. Note that in our sample, oxygen vacancies can easily be explained at high temperature by the loss of PbO weight during sintering[22-28], therefore we have:

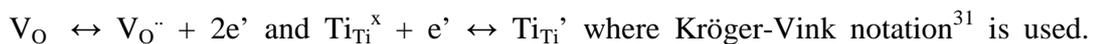

V$_O$ ↔ V$_O^{\cdot\cdot}$ + 2e' and Ti$_{Ti}^x$ + e' ↔ Ti$_{Ti}$' where Kröger-Vink notation[31] is used. Exciton transition interacting with the lattice could then occur between the localized hole V$_O^{\cdot\cdot}$ and the localized electron Ti$_{Ti}$'.

Very recently, Mochizuki et al. provide evidence for the crystal defect origin (oxygen defects and chemical heterogeneity in the surface region) of the photoluminescence of a $SrTiO_3$ single crystal[32] centered at 2.4 eV which could clearly be modified depending on the oxidizing-reducing atmosphere.

Two sets of Raman spectra obtained between 10 and 300 K with 514.5 and 647.1 nm excitation lines are shown on Fig. 2a and Fig. 2b respectively. The 514.5 nm spectra, Fig. 2a, are in clear accordance with those previously reported in the literature[8,11-14] where the temperature effect is clearly minor, which could be associated to the well-known static-dynamic disorder of PZT systems[5-9]. In contrast, the 647.1 nm Raman signal previously reported[28], Fig. 2b, exhibits major and continuous spectral changes as a function of the temperature; additionally, the changes are in close agreement with the transformation temperature previously reported by dielectric measurements[15-16], X-ray[17] and neutron diffraction[4]. This indicates that the same sample, which has only one phase transition sequence, was studied as a function of the temperature by two different laser energies but only the 647.1 nm probe light enabled these transformations to be detected. Based on this observation, we can conclude that the changes in temperature are clearly excitation line dependent. These changes are related therefore to the evolution of the resonance conditions and thus to the electronic levels involved as a function of the phase transition sequence in PZT materials. Interestingly, it transpires that the resonance behaviour using the 647.1 nm excitation line is structure-dependent what is an original phenomenon.

Corrected Stokes $(n(\omega,T)+1)/n(\omega,T)$ - anti-Stokes (S – AS) Raman spectra of $PbZr_{0.52}Ti_{0.48}O_3$ obtained at 150 K with the 514.5 nm excitation line and represented on the same abscissa, Fig. 3a, give clearly evidence of Bose statistics. Consequently, we did not observe any anomalously high local heating over the entire temperature range for spectra obtained with this radiation line. Corrected S - AS Raman spectra of PbZr$_{0.52}$Ti$_{0.48}$O$_3$ obtained at 200 K with the 647.1 nm excitation line, Fig. 3b, show a distinct phenomenon in which the S – AS ratio is not scaled as it is the case for example in the resonance process of carbon nanotubes[34]. This anomalous strong signal in the S - scattering can find its origin from a particular resonance behavior on emission, for which the phonon intensity would increase to a much greater extent under specific conditions on the S – rather than on the AS - side. This can explain the existence of a background scattering centered around 380 cm$^{-1}$ on the S – Raman spectra obtained as a function of the temperature, Fig. 4 (corrected ($n(\omega,T)+1$)). Thus to fit these spectra between 10 and 350 K, a quadratic baseline was first subtracted to take into account this peculiar background.

Beginning from 350 K, Fig. 4a, the Raman spectrum of PbZr$_{0.52}$Ti$_{0.48}$O$_3$ consists of several broad lines. Below 300 K, Fig. 5a, the Raman lines become more asymmetric, noticeably for the mode *1* and the region *2* and *3*, indicating the transformation to the monoclinic phase[28]. Upon decreasing temperature, a decrease of the line width occurs and at 200 K the modes *4* and *5* appear, Figs 4b-6a. These changes, which are represented on the spectrum obtained at 154 K, Fig. 5b, are clearly associated to the phase transition between the two monoclinic phases (*Cm-Cc*)[28]. The increase in the number of modes (Fig. 5d) at the phase transition is consistent with the greater number of modes predicted by group theory due to the doubling of the unit cell. Below 35 K, Fig. 5c, a brutal change in intensity of the entire spectrum occurs accompanied with the disappearance of mode *8* and the inversion in intensity ratio of mode *6-7*, Fig. 4b-5c-6b. These changes indicate an important modification to the local and/or lange range structure. Until now, no phase transition has been detected at such a temperature for this composition based on X-ray diffraction and neutron diffraction experiments[35-37]. Such behavior may instead result from a maximum in the phonon-intensity due to the resonance effect for a 1.92 eV laser energy, which may be reached in a discontinuous manner due to a change in the local structure around the titanium atoms in this disordered system. The temperature dependence of the Raman modes is summarized on the Fig. 5d. As described above, the transformation temperatures obtained from our data are in good agreement with those published previously.

## IV. CONCLUSIONS

In this study, we investigated the phase transition sequence of $PbZr_{0.52}Ti_{0.48}O_3$ as a function of the temperature by resonance Raman Spectroscopy. Using an excitation line of 647.1 nm with an energy close to a self-trapped level exciton energy deficient complex ($Ti_{Ti}'$-$V_O^{..}$) of PZT powder, we induced a resonance effect. As resonance Raman scattering spectroscopy is known to be a powerful method to detect the phonon modes which couple to an electronic system, this technique has clearly permitted us to distinguish the structural change of PZT in spite of its well-known static-dynamical disorder[5-9]. The resonance phenomenon resulted in an increase in the phonons intensities thereby enabling a clear dissociation of the different modes in the Raman spectra, at least at low temperature. Such an approach should also be very useful for other ferro/piezoelectric systems such as the important $PbMg_{1/3}Nb_{2/3}O_3$-$PbTiO_3$ (PMN-PT) and $PbZn_{1/3}Nb_{2/3}O_3$-$PbTiO_3$ (PZN-PT)-based materials with giant piezoelectric response as many perovskite oxides are known to display visible photoluminescence between 2 and 3 eV[20-21,38] very similar to that observed for PZT.

These ferroelectric-relaxor compounds also exhibit similar static and dynamic disorder in their solid solution and present also a MPB region what still lead the determination of their corresponding phase diagram to be a subject of debate. The possibility to investigate the temperature-composition behavior of these materials by Raman Resonance Spectroscopy would be of particular interest to clarify both dynamical properties and the phase transition sequence.

*Author to whom correspondence should be addressed. Email address: Jerome.Rouquette@uni-bayreuth.de

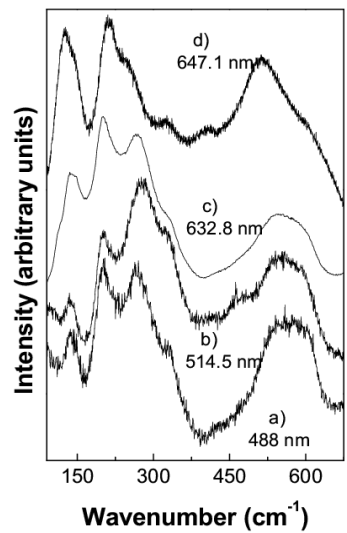

FIG. 1. Room temperature Raman spectra of PbZr$_{0.52}$Ti$_{0.48}$O$_3$ using as-followed excitation lines: a) 488 nm, b) 514.5 nm, c) 632.8 nm and d) 647.1 nm.

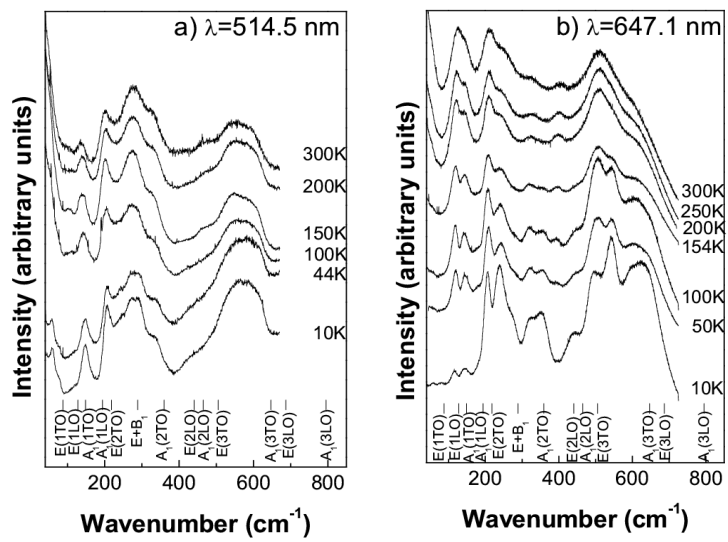

FIG. 2. Raman spectra obtained between 10 and 300 K with a) $\lambda_{laser}$ = 514.5, and b) $\lambda_{laser}$ = 647.1 nm. Positions of the Raman peaks at room temperature for $PbTiO_3$ are indicated (Ref. 33).

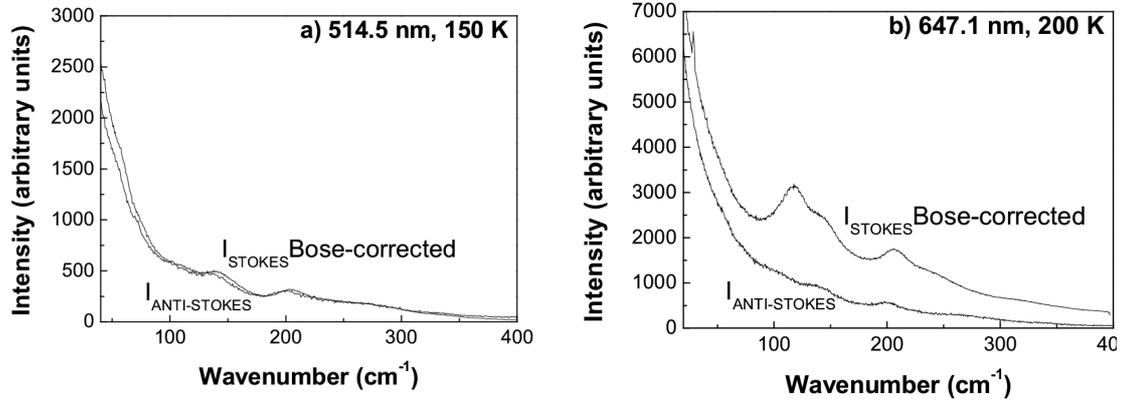

FIG. 3. Bose-corrected Stokes $(n(\omega,T)+1)/n(\omega,T)$-anti-Stokes Raman spectra of PbZr$_{0.52}$Ti$_{0.48}$O$_3$ represented on the same abscissa obtained at: a) 150 K with $\lambda_{laser}$ = 514.5 nm, and with $\lambda_{laser}$ = 647.1 nm at b) 200 K.

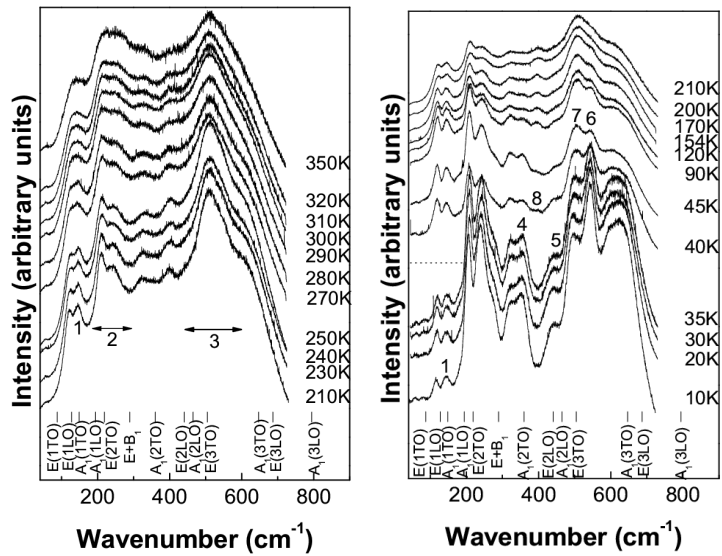

FIG. 4. Raman spectra obtained with the 647.1 nm excitation line: a) between 350 and 210 K, b) between 210 and 10 K. All spectra were corrected for Bose statistics ($n(\omega,T)+1$) and were also all arbitrary normalized to the mode *1* ($A_1$(1TO) for PbTiO$_3$) in order to take into account the intensity change as a function of temperature. Positions of the Raman peaks at room temperature for PbTiO$_3$ are indicated (Ref. 33).

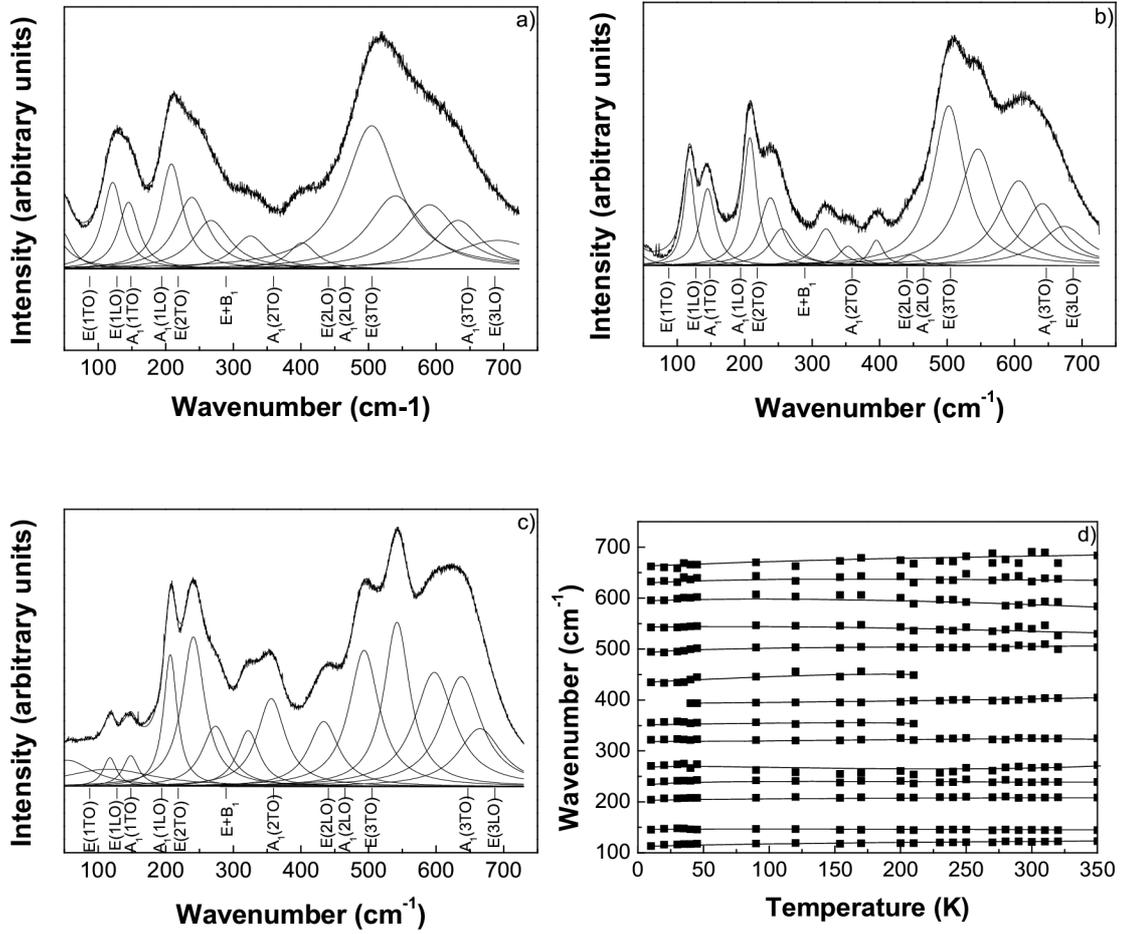

FIG. 5. Raman spectra obtained with the 647.1 nm excitation line measured at: a) 300 K, b) 154 K and c) 20 K (band at 180 cm$^{-1}$ was not fitted). All spectra were corrected for Bose statistics ($n(\omega,T)+1$) and were also all arbitrary normalized to the mode *1* (A$_1$(1TO) for PbTiO$_3$) in order to take into account the intensity change as a function of temperature. A quadratic baseline was then subtracted to the Raman line before fitting as-resulted spectra with a set of Lorentzian curves. Positions of the Raman peaks at room temperature for PbTiO$_3$ are indicated (Ref. 33) (d) temperature dependence of the Raman modes.

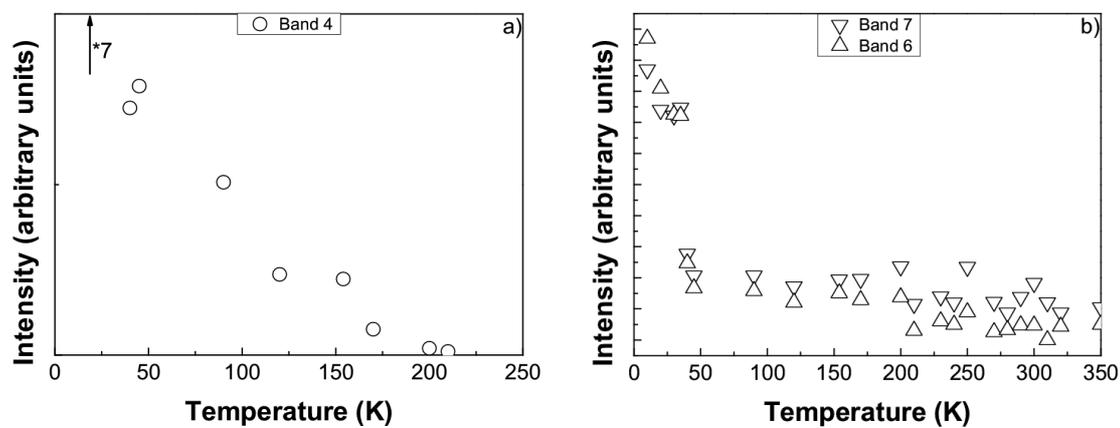

FIG. 6. Temperature dependence of the intensity of: a) band 4, b) band 6 and 7.